# ON DESIGNING FINITE TIME ITERATIVE LEARNING CONTROL BASED ON STEADY STATE FREQUENCY RESPONSE

Shuo Liu[1], Richard W. Longman[2] and Benjamas Panomruttanarug[3]

Iterative Learning Control (ILC) is useful in spacecraft application for repeated high precision scanning maneuvers. Repetitive Control (RC) produces effective active vibration isolation based on steady state frequency response. This paper considers the finite time ILC problem addressed using steady state frequency response, comparing two methods recently developed. One adapts for ILC the FIR filter design in RC that mimics the system's steady state frequency response inverse, creating a filter designed for all frequencies from zero to Nyquist. Adjustment of gains near the beginning of the matrix need to be made because FIR gains are truncated there. The other approach uses a circulant matrix obtained from the Toeplitz matrix of Markov parameters. It is shown to give steady state frequency response for the discrete frequencies that can be seen in the number of time steps in the ILC tracking problem. The main aim of the paper is to compare the performance of each approach, and compare the ease of use. The inverse of the frequency response is used in RC to avoid the usually unstable inverse of the discrete time transfer function. Here the use in ILC has the same property. The performance and the robustness to model parameter error do not seem to have any significant difference. The main distinction is that the adjustment by steepest descent of a small number of gains in the Learning Gain Matrix is easier done in the FIR approach.

## INTRODUCTION

Classical control system design is divided into two main branches, design in the time domain based on differential equation models, or design in the frequency domain using Bode plots and Nyquist stability criterion, etc. Both approaches can be used in digital control system design. It is interesting that frequency response methods which seem more abstract, predate such time domain techniques as root locus plots. This is perhaps the result of the fact that early development came from Bell Telephone Laboratories whose focus was on transmission and boosting of sound signals.

Iterative learning control (ILC), see References [1-3], and repetitive control (RC), see references [4-7], are two types of control that learn from experience performing a specific command. RC is useful in spacecraft applications to create active vibration isolation mounts of fine pointing equipment. ILC considers finite time tracking maneuvers for repeated sensor scanning, aiming to have precise sensor pointing as a function of time. ILC differs from RC in the fact that it does finite

[1] Department of Mechanical Engineering, Boston University, 110 Cummington Mall, Boston, MA 02215
[2] Professor of Mechanical Engineering, Department of Mechanical Engineering, Columbia University, MC4703, 500 West 120th Street, New York, NY 10027
[3] Department of Control System and Instrumentation Engineering, King Mongkut's University of Technology Thonburi, Bangmod Tungkru, Bangkok, Thailand 10140

time maneuvers repeatedly, each time starting from the same starting point. References [6-7] present experimental results testing RC on a floating satellite testbed in a laboratory with the attitude control system running.

Frequency response is a natural approach to designing RC systems whose objective is to make a control system converge to zero tracking error following a periodic command, or converge to zero tracking error following a constant command in the presence of a periodic disturbance. Convergence can be obtained as time tends to infinity allowing frequency response thinking to apply, Reference [4]. Frequency response is by definition steady state so that it applies after all transients have become negligible. ILC wants zero error throughout the whole-time interval following a finite time command, which includes any transients occurring from the start of each repeat of the maneuver. Hence, ILC designs are normally done in the time domain.

This paper studies the use of steady state frequency response approaches to design ILC laws that apply to the finite time signals in ILC problems, and includes the transients that appear in each run or iteration in ILC.

**POTENTIAL BENEFITS OF DESIGNING ILC CONTROL LAWS BASED ON FREQUENCY RESPONSE**

This paper examines the use of the inverse of the steady state system frequency response as the learning control gain in ILC. Some potential benefits include the following:

(1) Considering the general field of control theory, being able to use frequency response as the basis for ILC design fills a hole in the theory.
(2) ILC is an inverse problem, find the input that produces the desired output. It must be digital because it must store data from the previous run. We model the situation when a continuous time system is fed through a zero order hold, and the output examined at sample times. According to Reference [8], one expects this inverse problem to have an unstable inverse for any system with a pole excess of 3 or more. Therefore, one can say that the usual ILC problem is not well posed.
(3) While the inverse of the transfer function is likely unstable, the inverse of the steady state frequency response does not have this problem.
(4) There are several approaches in the literature to make a stable inverse. The common approach as in Reference [9] requires pre and post actuation and zero initial and final conditions that are zero. It is not appropriate for ILC. References [10,11] give a different approach to stabilizing the ILC inverse problem, by not asking for zero tracking error at least for one or a few initial time steps.
(5) Using the steady state frequency response inverse offers an alternative to these stabilization methods. The instability is characterized by an unstable growth behavior at Nyquist frequency. The steady state frequency response inverse at this frequency is likely unable to capture this growing signal, and hence should avoid the unstable behavior. And, one can of course also remove Nyquist from the inverse frequency response to address the unstable inverse issue.
(6) Use of steady state frequency response for controller design offers the potential of simply using frequency response test data as the basis for the ILC design, which avoids the process of developing a model from the data.
(7) Evaluating robustness of control systems is often done based on model uncertainty given as a function of frequency. In this case one can adjust the gain as a function of frequency to improve robustness. There are also potential advantages in handling unmodeled high frequency dynamics with a cut off of the learning, one must be careful to examine the behavior when the cutoff is not perfect.

# TWO APPROACHES TO CREATING A FINITE TIME VERSION OF A SYSTEM'S STEADY STATE FREQUENCY RESPONSE

The use of a Finite Impulse Response model of the system steady state frequency response inverse was shown to be very effective in Repetitive Control in Reference [4]. Reference [12] examined the use and issues involved when one tries to use the same FIR model in ILC. A recent Reference [13] investigated the use of the circulant matrix of system Markov Parameter in place of the FIR filter, as suggested by References [14,15]. Both approaches are further investigated here, with a major objective to examine the advantages and disadvantages of each approach.

*Approach 1:* Given the z-transfer function $G(z)$ of a feedback control system, one computes the magnitude and phase response by finding the magnitude and phase of $G(e^{i\omega T})$ where $\omega$ is the radian frequency from zero to Nyquist, and $T$ is the sample time interval. An FIR filter is designed as a curve fit to the inverse of this frequency response, sampled at an appropriate number of frequencies between zero and Nyquist. Some properties and issues for this approach are as follows.

(1) This approach is based on a curve fit and may not be perfect. But experience in Reference [4] using only 12 gains in the filter applied to the third order system used later here for numerical studies, captured the frequency response to about 3 digit accuracy. Note that this result says that one can compute the desired result using just a simple linear combination of 12 error measurements from the previous period for RC, or previous run for ILC.
(2) Reference [12], when investigating a 51 time step long desired trajectory filled the square learning gain matrix using 51 gains. This completely fills the middle row of the matrix, but leaves two triangular sets of zeros, one in the lower left corner of the matrix, and one in the upper right corner of the matrix. We do comparisons of the ILC approaches making use of this matrix.
(3) For completeness, we also study the influence of computing enough gains in the FIR filter to fill in those zero entries in the lower left and upper right corners.
(4) The FIR filter is non-causal. In the 12 gain result in item (1) above there are 6 gains forward in time, and 5 backward in time (the lack of symmetry is based on the one time step delay between input and output due to the zero order hold). One wants to apply this filter to every time step of the error in the previous run, but in the upper left corner, for the first few time steps there are missing data points for the filter terms that relate to error at time steps from before the start of the ILC run. One expects serious errors in the computation of the intended inverse frequency response as a result of these truncated gains. The same is true at the bottom right corner of the matrix, where forward time step data is missing near the end of the trajectory. Experience says the former problem must be addressed. The latter problem may not cause serious trouble.
(5) A serious consideration in using steady state frequency response based ILC design, is that the start of every run will have initial transients. Some gains in the upper left corner of the learning gain matrix, that apply to time steps at the beginning of the trajectory during transients, need to be adjusted. It is perhaps surprising, in simulations in Reference [12] adjusting just a 2 by 2 set of gains is sufficient to result in a stable ILC control system. This number seems unrelated to the number of time steps in the system settling time for transients.
(6) Adjustment of these few gains is done by a steepest descent approach of the maximum singular value of the matrix giving the update of the error from run to run. This addresses both the issue in item (4) and the issue in item (5) above. Which gains to adjust is initially studied by a sensitivity analysis indicating which gains have the most influence on the largest singular value.
(7) One may choose to not ask for zero error at the first, or first few time steps, in order to make the ILC problem well posed, i.e. to have a stable inverse solution.

After addressing these issues, a very effective ILC law can be designed, and convergence to zero tracking error is much quicker than in the typical time domain ILC laws.

*Approach 2:* The second approach makes use of the circulant matrix of Markov Parameters, and is studied in Reference [13]. The big apparent advantage of this matrix is that when given an input of a frequency that can be seen in the number of time steps of the desired trajectory, it actually gives the steady state frequency response. No curve fitting. This matrix is obtained from the lower triangular Toeplitz matrix whose first column is the unit pulse response history per time step of the system. One obtains the second column by moving each entry down, and then taking the entry that went out of the bottom of the matrix and inserting at the vacated top entry. Continuing in this way produced the circulant matrix of interest here. Many of the same issues in Approach 1 are also present in Approach 2. Some properties and issues in using the inverse of this matrix are as follows.

(1) Of course, it is very attractive to have a matrix that is obtained analytically that gives the steady state frequency response.
(2) Unfortunately, it only gives the steady state frequency response for the discrete set of frequencies that can be seen in the number of time steps in the desired trajectory. Every frequency must be periodic with the number of time steps in the trajectory. So the frequency components are expected to come back to the original value at the end of the trajectory.
(3) It is important to examine what the ILC law does to signals that do not return to their original values at the end of the trajectory. This property would be expected to apply to essentially all iterations of the ILC, and likely it applies to the final desired trajectory as well.
(4) Again, one must adjust some gains in the ILC law in the top left corner that apply to the first steps of the trajectory. Again, sensitivity analysis is used to find which gains to adjust in order to have the most influence on the maximum singular value that is used as a criterion for monotonic convergence. The adjustment is again done by steepest descent. Numerical experiments given below suggest adjusting more gains than in Approach 1.
(5) Numerical experience with Approach 1 suggested that there was no need to adjust gains near the end of the trajectory. But Approach 2 wants all signals to be periodic coming back to the starting point at the end of the trajectory, and it is found that adjusting a set of gains near the top right corner of the learning gain matrix is necessary for this approach.

We will see that design by this approach can also result in effective ILC designs.

**BASIC FORMULATIN OF ILC**

The general formulation of ILC as developed in Reference [16] is presented in this section. Consider a single-input single-output system in state space form

$$x(k+1) = Ax(k) + Bu(k) \quad k = 0,1,2,\ldots,N-1 \tag{1}$$
$$y(k) = Cx(k) \quad k = 1,2,\ldots,N$$

Denote the desired $N$ time step output as $\underline{y}^*$, the $N$ time step output at iteration $j$ as $\underline{y}_j$, the corresponding command that produces the output at iteration $j$ as $\underline{u}_j$, and the tracking error history at iteration $j$ as $\underline{e}_j = \underline{y}^* - \underline{y}_j$. The command is adjusted iteratively in order to generate zero tracking error as the repetition number $j$ tends to infinity. Underbar indicates a column vector of the history of the variable during an iteration, or the history of a quantity in the initial conditions, as given by

$$\underline{y}^* = [y^*(1), y^*(2), \ldots, y^*(N)]^T \tag{2}$$
$$\underline{y}_j = [y_j(1), y_j(2), \ldots, y_j(N)]^T$$
$$\underline{e}_j = [e_j(1), e_j(2), \ldots, e_j(N)]^T$$
$$\underline{u}_j = [u_j(0), u_j(1), \ldots, u_j(N-1)]^T$$

Note that there is one time step delay from input history to output history, corresponding to a zero order hold input-output delay, and similarly for the tracking error history. This applies when Equation (1) is the closed loop version of a continuous time feedback system with the command coming through a zero order hold and updated each sample time. Simple modifications can be made for a digital feedback control system typically with a two time step delay, one for the controller, and one for the plant fed by a zero order hold.

The general linear ILC law is given as

$$\underline{u}_{j+1} = \underline{u}_j + L\underline{e}_j \tag{3}$$

where the learning gain matrix $L$ has $N \times N$ learning gains chosen by the designer. This paper will pick this matrix based on the reciprocal of the steady state frequency response using either Approach 1 or Approach 2. The repetition domain system model can be created by recursively applying Equation (1), writing the convolution sum solution for each time step $k$ of an iteration $j$ as

$$y_j(k) = CA^k x(0) + \sum_{i=0}^{N-1} CA^{k-i-1} B u_j(i) \tag{4}$$

and packaging the result in matrix form to create an input-output model of the system for each iteration $j$

$$\underline{y}_j = P\underline{u}_j + Ox(0) \tag{5}$$

Matrix $P$ is a lower triangular Toeplitz matrix containing the Markov parameters, or the unit pulse response history of the system. Vector $O$ is an $N \times 1$ step observability column vector

$$P = \begin{bmatrix} CB & 0 & \cdots & 0 \\ CAB & CB & \cdots & \vdots \\ \vdots & \vdots & \ddots & 0 \\ CA^{N-1}B & CA^{N-2}B & \cdots & CB \end{bmatrix} \qquad O = \begin{bmatrix} CA \\ CA^2 \\ \vdots \\ CA^N \end{bmatrix} \tag{6}$$

The ILC problem assumes that the system is reset to the same initial condition before each iteration, so Equation (5) implies $\underline{y}_j - \underline{y}_{j-1} = P(\underline{u}_j - \underline{u}_{j-1})$. Based on this equation and Eq. (3), one can compute the error propagation matrix from one iteration to the next, and from error in the first run to current error in the $j^{th}$ run

$$\underline{e}_j = (I - PL)\underline{e}_{j-1} \tag{7}$$
$$\underline{e}_j = (I - PL)^j \underline{e}_0$$

One can show that the error $\underline{e}_j$ will converge to zero error as $j \to \infty$ for all possible initial condition runs $\underline{e}_0$, if and only if the spectral radius of matrix $I - PL$ is less than one, i.e. all eigenvalues are less than one in magnitude

$$|\lambda_i(I - PL)| < 1 \tag{8}$$

Some ILC laws exhibit bad transients during the learning process, Reference [17]. The following condition is a sufficient condition for monotonic decay and convergence to zero error of the Euclidean norm of the error vector as $j \to \infty$. Compute the singular value decomposition $I - PL = U_1 S_1 V_1^T$. The error history vector in the $j^{th}$ run is multiplied by the transpose of the left singular vector matrix as $U_1^T e_j = S_1 V_1^T e_{j-1}$. Since $U_1^T$ and $V_1^T$ are unitary, the products do not change the Euclidean norm of the error. Then asking that the maximum singular value of this matrix given in matrix $S_1$ be less than one

$$\sigma_i(I - PL) < 1 \tag{9}$$

guarantees monotonic decay. Both to ensure well behaved transients, as well as for computational convenience, it is this singular value condition that we use when adjusting gains by steepest descent in the learning matrices of Approaches 1 and 2 to produce convergent ILC laws based on the inverse of the steady state system frequency response.

## DESIGN OF THE FIR INVERSE FREQUENCY RESPONSE FILTER IN APPROACH 1

The FIR filter created in Reference [4] for RC design, uses a filter of the following form

$$F(z) = a_1 z^{m-1} + a_2 z^{m-2} + \cdots + a_m z^0 + \cdots + a_n z^{-(n-m)} \tag{10}$$

The $z^0$ term corresponds to the time step associated with the filtered results, and there are $m - 1$ steps forward in time, and $(n - m)$ steps backward in time. The values of $m$ and $n$ are important when the filter only has a small number of terms as in the 12 gains discussed in Approach 1, but when a large number of gains is used, the choice is no longer critical.

The gains are chosen to minimize the cost function

$$J = \sum_{j=0}^{N} \left[1 - G(e^{iw_j T}) F(e^{iw_j T})\right] \left[1 - G(e^{iw_j T}) F(e^{iw_j T})\right]^* \tag{11}$$

which tries to make the filter frequency response times the system frequency response look like one. The summation is taken over a sufficiently dense set of frequency samples $\omega_j$ between zero and Nyquist (computations are made using every degree from 0 to 179). This is a quadratic cost with respect to the coefficients in Equation (10), and it leads to a linear set of equations $A\chi = b$ to solve for the coefficients, where

$$A = \sum_{j=0}^{N} M_G(\omega_j)^2 \begin{bmatrix} 1 & \cos(\omega_j T) & \cdots & \cos((n-1)\omega_j T) \\ \cos(\omega_j T) & 1 & \cdots & \cos((n-2)\omega_j T) \\ \vdots & \vdots & \ddots & \vdots \\ \cos((n-1)\omega_j T) & \cos((n-2)\omega_j T) & \cdots & 1 \end{bmatrix}$$

$$\chi = [a_1, a_2, \ldots, a_n]^T$$

$$b = \sum_{j=0}^{N} M_G(\omega_j) \begin{bmatrix} \cos((m-1)\omega_j T + \theta_G(\omega_j)) \\ \cos((m-2)\omega_j T + \theta_G(\omega_j)) \\ \vdots \\ \cos((m-n)\omega_j T + \theta_G(\omega_j)) \end{bmatrix} \tag{12}$$

The $M_G$ and $\theta_G$ are the magnitude and phase of $G(e^{i\omega_j T})$. These gains are used to fill the Learning Control Matrix $L$. This matrix based on the FIR filter will be denoted by the letter $F$ without the argument $z$. There is a one time step delay through the system, so the coefficient $a_m$ corresponding to the current time step should fill the first sub-diagonal, i.e., the diagonal immediately below the main diagonal in the matrix. When the number of time steps in the desired trajectory is 51 or 101, then we choose to find 51 or 101 gains, and they fill the middle row of the matrix. Rows above that will be missing one or more of the backward steps that no longer fit into the matrix, and rows below that will be missing one or more of the forward terms. Also considered for comparison purposes is the computation of enough gains to fill all rows of the matrix.

Often no attempt will be made to converge to zero error of the first time step of the desired trajectory. This applies to the case when there is one zero outside the unit circle as in a pole excess of 3 (more if the pole excess is larger). In that case, a subscript "1" is introduced, $F_1$, indicating that the first column of $F$ has been removed. Correspondingly, the first row of $P$ is removed indicated by $P_1$. When the matrix $F$ has gains filling the whole matrix, an $f$ subscript is added, $F_{f1}$. Starting from Equation (7), we can study convergence properties by examining the maximum singular value of the coefficient matrices in the following equations. Note that the dimension of the square matrix $(I - PL)$ has been reduced by one in the process, and the error vector $\underline{e}_j$ dimension has also been reduced by eliminating consideration of the first time step.

$$\underline{e}_j = (I - P_1 F_1)\underline{e}_{j-1} \tag{13a}$$
$$\underline{e}_j = (I - P_1 F_{f1})\underline{e}_{j-1} \tag{13b}$$

Consider the following system as an example, produced by discretizing the continuous time transfer function

$$G(s) = \left(\frac{a}{s+a}\right)\left(\frac{\omega_0^2}{s^2 + 2\xi\omega_0 s + \omega_0^2}\right)$$

$$a = 8.8;\ \omega_0 = 37;\ \xi = 0.5 \tag{14}$$

fed by a zero order hold, sampled at 100 Hz, with 101 time steps in $u(k)$, and the result converted to state variable form. Because the first time step is not considered in the learning process, the matrices are 100 by 100. The singular values for each approach are presented in Table 1 and Table 2.

Table 1 First and Last Six Singular Values of $(I - P_1 F_1)$ from Large to Small

| Order | $\sigma_1$ | $\sigma_2$ | $\sigma_3$ | $\sigma_4$ | $\sigma_5$ | $\sigma_6$ |
|---|---|---|---|---|---|---|
| Singular value | 17.9361 | $8.5440e^{-10}$ | $6.3403e^{-10}$ | $6.1560e^{-10}$ | $5.1517e^{-10}$ | $4.7879e^{-10}$ |
| Order | $\sigma_{95}$ | $\sigma_{96}$ | $\sigma_{97}$ | $\sigma_{98}$ | $\sigma_{99}$ | $\sigma_{100}$ |
| Singular value | $2.1149e^{-12}$ | $1.8130e^{-12}$ | $1.4568e^{-12}$ | $5.7490e^{-13}$ | $2.7823e^{-13}$ | $2.0884e^{-14}$ |

Table 2 First and Last Six Singular Values of $(I - P_1 F_{f1})$ from Large to Small

| Order | $\sigma_1$ | $\sigma_2$ | $\sigma_3$ | $\sigma_4$ | $\sigma_5$ | $\sigma_6$ |
|---|---|---|---|---|---|---|
| Singular value | 17.9361 | $7.9210e^{-10}$ | $4.8413e^{-10}$ | $4.6789e^{-10}$ | $3.8082e^{-10}$ | $3.6662e^{-10}$ |
| Order | $\sigma_{95}$ | $\sigma_{96}$ | $\sigma_{97}$ | $\sigma_{98}$ | $\sigma_{99}$ | $\sigma_{100}$ |
| Singular value | $2.7032e^{-12}$ | $2.0266e^{-12}$ | $1.9690e^{-12}$ | $1.3494e^{-12}$ | $7.0231e^{-13}$ | $2.5736e^{-13}$ |

There are a few conclusions. Something must be done to reduce the maximum singular value. All other singular values have fast learning from one iteration to the next using Approach 1. Also, the change in performance by filling up all entries in the learning gain matrix as $F_{f1}$ instead of $F_1$ is negligible.

## THE CIRCULANT MATRIX OF APPROACH 2

The circulant matrix in Approach 2 is

$$P_c = \begin{bmatrix} CB & CA^{N-1}B & \cdots & CA^2B & CAB \\ CAB & CB & \cdots & CA^3B & CA^2B \\ \vdots & \vdots & \ddots & \vdots & \vdots \\ CA^{N-2}B & CA^{N-3}B & \cdots & CB & CA^{N-1}B \\ CA^{N-1}B & CA^{N-2}B & \cdots & CAB & CB \end{bmatrix} \qquad (15)$$

It is obtained from matrix $P$ of Equation (6) by creating each column from the previous column by moving all entries down one, and the entry that went out of the matrix at the bottom is put in at the top. Reference [14] suggested the use of this matrix as a cutoff filter in ILC for robustification, and supplied a proof that it produced the steady state response of the $A, B, C$ system. Reference [15] studies in more detail the use of a circulant matrix as a cutoff filter, and supplies a different proof. Here we investigate its use for a different purpose, using it to represent the inverse of the frequency response as a Learning Control Gain Matrix, in Equation (3).

The property of $P_c$ that it gives the steady state frequency response in a compact matrix form of the dimension of our ILC problem, is very attractive. However, it only supplies this steady state result for the frequencies one can see in the $N$ time steps of data available in each run. If $N$ is even then one can see DC (zero frequency), Nyquist frequency, and $(N-2)/2$ other frequencies. If $N$ is odd, then one can see DC, and $(N-1)/2$ other frequencies, without Nyquist. An important issue is to investigate what happens to frequencies between these discrete frequencies.

Figures 1 and 2 examine how the various matrices respond to a sine wave input and a cosine wave input. Figure 1 presents the RMS deviation as a function frequency of the frequency response of system model $P$ compared to the frequency response of the system $z$-transfer function associated with the third order system of Equation (14), and the same for the circulant matrix $P_c$ and 10 times size extended circulant matrix $P_{ec}$. The outputs $y(k)$ from $\underline{y} = P\underline{u}$ and $\underline{y} = P_c\underline{u}$ where the $\underline{u}$ is the column vector of the 100 samples of $u(k) = \sin(\omega kT)$ (left plot) and $u(k) = \cos(\omega kT)$ (right plot) and the output $y(k)$ from $\underline{y} = P_{ec}\underline{u}$ where the $\underline{u}$ is the column vector of the 1000 samples of $u(k) = \sin(\omega kT)$ (left plot) and $u(k) = \cos(\omega kT)$ (right plot). As predicted, the frequency response of the circulant matrix $P_c$ has zero error at the frequencies that one can see in $N$ time steps, which are the turning points of the vibrating curve plotted on the horizontal axis while the frequency response of the extended circulant matrix $P_{ec}$ has zero error at the frequencies that 10 times finer than frequencies that one can see in $N$ time steps by circulant matrix $P_c$. Figure 2 gives a detailed view that shows how the frequency response of the circulant matrix behaves between these frequencies, after extending the desired trajectory 10 times. The error in the extended trajectory is reduced below that from the $P$ matrix, but is still substantial at the frequencies that one cannot see. The resulting behavior in the context of application to ILC is investigated here.

A significant factor contributing to these between frequency errors is Gibbs phenomenon. When the start point and the endpoint of the trajectory do not match, the Gibbs phenomenon reacts as if there is a cliff discontinuity at the start and the end, and produces oscillations at the start and the end of the trajectory (Reference [15]).

One way to address error at frequencies between the integer frequencies, is to take the original signal that is 100 time steps long, simply repeat that signal after the end of the original 100 steps, and keep repeating it to make a signal that is 1000 time steps long. Then use the first 100 time steps of the result. This makes the number of frequencies that can be seen in the given data points be about 10 times as many. As a result, the number of frequencies where the circulant matrix gives the correct steady state frequency response, becomes 10 times as many. The corresponding results produce a plot that lies on top of the zero error line as seen to graphical accuracy. The approach clearly can work very well. But of course, to use it one pays a price in increased computation. After agreeing to not try to learn the first time step, one can study the convergence behavior using the circulant learning gain matrix $P_{c1}^{-1}$ obtained by taking the inverse of $P_c$ and then deleting the first column, or the extended version with learning gain matrix denoted by $P_{ec1}^{-1}$ by studying the singular values of the coefficient metrices in

$$\underline{e}_j = (I - P_1 P_{c1}^{-1})\underline{e}_{j-1} \tag{16a}$$
$$\underline{e}_j = (I - P_1 P_{ec1}^{-1})\underline{e}_{j-1} \tag{16b}$$

The singular values for these coefficient matrices associated with using the inverse circulant matrix to form the learning gain matrix, are shown in Tables 3 and 4. Each singular value indicates the decay of error for the associated singular vector error component. Note that the original size of $P$ here is 101 by 101 so $P_1$ is 100 by 101, $P_{c1}^{-1}$ is 101 by 100, $P_{e1}$ is 1009 by 1010, $P_{ec1}^{-1}$ is 1010 by 1009.

**Table 3 First and Last Six Singular Values of $(I - P_1 P_{c1}^{-1})$ from Large to Small**

| Order | $\sigma_1$ | $\sigma_2$ | $\sigma_3$ | $\sigma_4$ | $\sigma_5$ | $\sigma_6$ |
|---|---|---|---|---|---|---|
| Singular value | 84.2474 | 1.7244 | 0.2341 | 0.0146 | 0.0146 | 0.0145 |
| Order | $\sigma_{95}$ | $\sigma_{96}$ | $\sigma_{97}$ | $\sigma_{98}$ | $\sigma_{99}$ | $\sigma_{100}$ |
| Singular value | $1.5341e^{-4}$ | $1.4900e^{-4}$ | $1.4864e^{-4}$ | $2.4385e^{-7}$ | $6.9588e^{-8}$ | $3.5668e^{-14}$ |

Table 4 First and Last Six Singular Values of $(I - P_{e1}P_{ec1}^{-1})$ from Large to Small

| Order | $\sigma_1$ | $\sigma_2$ | $\sigma_3$ | $\sigma_4$ | $\sigma_5$ | $\sigma_6$ |
|---|---|---|---|---|---|---|
| Singular value | 85.2206 | 1.7435 | 0.2388 | $1.8838e^{-12}$ | $1.7151e^{-12}$ | $1.5848e^{-12}$ |
| Order | $\sigma_{1004}$ | $\sigma_{1005}$ | $\sigma_{1006}$ | $\lambda_{1007}$ | $\lambda_{1008}$ | $\lambda_{1009}$ |
| Singular value | $1.0543e^{-15}$ | $1.0022e^{-15}$ | $8.7170e^{-16}$ | $6.1653e^{-16}$ | $2.7520e^{-16}$ | $1.1262e^{-16}$ |

Again it is necessary to reduce the maximum singular value, and this time it is substantially larger in Approach 2 than in Approach 1 as indicated in Tables 1 and 2. Also, this time there is one more singular value that is larger than unity.

**TUNING GAINS TO ADDRESS THE SIZE OF THE MAXIMUM SINGULAR VALUE**

ILC laws from Approach 1 and Approach 2 are based on steady state frequency response. But the ILC problem is finite time, and each run has transients at the start. One might expect that perhaps gains throughout the settling time of the transients might have to be adjusted for that reason. It is perhaps surprising that this is not the case. Another issue is that in Approach 1, near the beginning of the learning gain matrix, terms in the FIR frequency response fit must be truncated because they ask for data from before the ILC starts. Truncating some of the gains one would expect to have a strong effect on whether the FIR filter for those early time steps produces anything like the intended inverse frequency response. This is a second reason to expect to need to adjust gains in the upper left corner of the matrix. Based on these comments, we should not be surprised that the singular value for either Approaches 1 and 2 have one or more singular values larger that unity.

As described in Equations (8) and (9), the actual stability boundary is based on the spectral radius, and the singular value condition is a sufficient condition with the added property of monotonic decay of the Euclidean norm of the error. We choose to work with the maximum singular value instead of spectral radius. The monotonic decay property is desirable, mathematically singular values are much easier to deal with, and also we can accomplish the stabilization we seek using the criterion. The derivative of a singular value with respect to a matrix parameter is based on the derivative of an eigenvalue of a symmetric matrix which has simple formulas, see for example Reference [18].

Both Reference [12] and Reference [13] start with a sensitivity analysis of many or all entries in the corresponding Approach 1 or Approach 2 matrix, to make an initial assessment of which gains in the matrix are most effective at reducing the maximum singular value. The study for Approach 1 in Reference [12], investigated adjusting all entries in the first row or two, first column or two, and settled on adjusting a 2 by 2 block of the learning matrix in the upper left corner after the first column has been deleted. The resulting matrix is denoted by $F_{o1}$. We note that the FIR filter has truncated gains in the forward direction in the bottom right corner of the matrix, but unlike the upper left corner, the lower right corner truncations of gains seems to have little influence on the maximum singular value. No adjustment seems to be necessary there.

Reference [13] does a similar sensitivity analysis and concluded that more gains need to be adjusted in Approach 2, the computations here are made adjusting a 5 by 5 block in the upper left corner. And this time there is a need to adjust gains in the upper right corner, again, a 5 by 5 block is adjusted. The resulting matrix after inverting, deleting and optimizing is denoted by $P_{oc1}^{-1}$. The fact that the circulant wants to consider only signals that have the same start and end time is likely related to this need.

Adjustment of the chosen gains is made by following the steepest descent direction for the set of gains involved. We seek to make a fair comparison of Approach 1 design versus Approach 2 design. We adjust gains in both approaches so that each approach has the same maximum singular value of 0.55. Table 5 presents the singular values of $(I - P_1 F_{o1})$ after the optimization of 4 gains, and Table 6 presents the singular values of $(I - P_1 P_{oc1}^{-1})$ after the optimization of two blocks of 5 by 5 gains. Obviously, the optimization of the associated small number of gains in the full frequency response matrix can be very effective at reducing the maximum singular value sufficiently to make stable ILC based on the inverse of the steady state frequency response. And it is not only a stable ILC but one with monotonic decay of the Euclidean norm of the error, and there is fast error reduction each iteration, arbitrarily chosen to be a factor of 0.55 or more reduction each iteration.

Table 5 First and Last Six Singular Values of $(I - P_1 F_{o1})$ from Large to Small

| Order | $\sigma_1$ | $\sigma_2$ | $\sigma_3$ | $\sigma_4$ | $\sigma_5$ | $\sigma_6$ |
|---|---|---|---|---|---|---|
| Singular value | 0.5499 | 0.3080 | $3.9277e^{-12}$ | $3.8888e^{-12}$ | $3.8843e^{-12}$ | $3.8173e^{-12}$ |
| Order | $\sigma_{45}$ | $\sigma_{46}$ | $\sigma_{47}$ | $\sigma_{48}$ | $\sigma_{49}$ | $\sigma_{50}$ |
| Singular value | $1.9355e^{-14}$ | $1.4675e^{-14}$ | $1.4245e^{-14}$ | $9.9468e^{-15}$ | $1.6135e^{-15}$ | $5.1347e^{-16}$ |

Table 6 First and Last Six Singular Values of $(I - P_1 P_{oc1}^{-1})$ from Large to Small

| Order | $\sigma_1$ | $\sigma_2$ | $\sigma_3$ | $\sigma_4$ | $\sigma_5$ | $\sigma_6$ |
|---|---|---|---|---|---|---|
| Singular value | 0.5497 | 0.3291 | 0.0329 | 0.0159 | 0.0034 | 0.0034 |
| Order | $\sigma_{45}$ | $\sigma_{46}$ | $\sigma_{47}$ | $\sigma_{48}$ | $\sigma_{49}$ | $\sigma_{50}$ |
| Singular value | $1.3957e^{-4}$ | $1.3873e^{-4}$ | $1.1893e^{-4}$ | $2.7479e^{-6}$ | $2.8247e^{-7}$ | $4.2294e^{-8}$ |

## COMPARISON OF THE LEARNING PROGRESS IN ITERATIONS BY THE TWO APPROACHES

**Comparison of the RMS Error as a function of ILC Iterations:** Now that both Approach 1 and Approach 2 have a convergent ILC laws with matched maximum singular value, we compare the behavior of the learning processes. Using the matched maximum singular values for the two approaches, Figure 3 examine the Root Mean Square (RMS) error as a function of iteration number. The desired output is a 5[th] order polynomial that starts from rest at time zero, has continuous first and second derivatives across time zero, but requires a step input at zero to match the third derivative. There appears to be little difference of importance in the two plots.

**Examining the First Few ILC Iterations:** Figures 4 and 5 compare the first few iterations by both methods. The sample rate is 50 Hz and the initial run always applies the desired trajectory to the system, which for these plots is given by $y^*(kT) = \pi(1 - \cos(\omega kT))^2$. A 2 second period is chosen, and the desired trajectory is limited to the first second, and a 50Hz sample rate is used. This trajectory has a particularly smooth startup from rest before time zero, with the function, its first, second, and third derivatives are continuous at zero, for the third order system of Equation (14). Figure 4 plots the desired trajectory and the output after one iteration for learning, left plot for circulant, right plot for FIR approach. Note that the left plot exhibits some deviations about the desired trajectory for a short time at the beginning of the trajectory, but the FIR approach in the right plot gives better performance no error to plotting accuracy. We will use the word "wiggles" to describe these deviations as they often deviate in one direction and then the other. Figure 5 gives the corresponding results for the 3$^{rd}$ iteration, and the wiggles at the beginning for the circulant approach seem to be gone.

Figure 6 considers a sinusoidal desired trajectory $y^*(kT) = \sin(\omega kT)$. Note that considering this as a continuous time desired output, when substituted into the differential equation associated with Equation (14) requires that the input have the derivative of a Dirac delta function (a unit doublet) which no physical system could perform. Making it into a discrete time problem allows one to get zero error, but one must expect some discrete approximation of the wild input behavior. In Figure 6 after one iteration for learning one observes some rather small wiggling behavior near the beginning for both approaches. After 3 iterations these wiggles appear to be gone.

Each of the desired trajectories considered so far have different starting and ending points. The Gibbs phenomenon was discussed above, that suggests that perhaps one should expect some wiggles occurring near the start and the end of the trajectory when these points are different. Figure 7 examines a case of when the start and the endpoints coincide, using the same desired trajectory as in Figures 4 and 5, but with the trajectory corresponding to one period of one second. This means that the desired trajectory only contains frequencies that are periodic with the given number of time steps, so that the circulant matrix can perfectly represent each of these components. Note that even if the desired trajectory has an integer number of periods in the given time interval, that property applies to the converged trajectory, and the trajectories during the learning process are very unlikely to have this property. One observes wiggles near the beginning in the first iteration for the circulant ILC law on the left of the figure, but no wiggles visible on the right FIR design. Again, wiggles are gone by iteration 3.

## EXAMINING HOW ILC BASED ON STEADY STATE FREQUENCY RESPONSE CAN LEARN WHEN THE WHOLE TRAJECTORY IS WITHIN THE SYSTEM SETTLING TIME

These learning laws are based on the reciprocal of the steady state frequency response, so one might expect them to work on portions of the trajectory that can be considered to have reached steady state, i.e. after the influence of initial conditions have reached a negligible contribution to the observed output. One can gage this time based on the settling time of the system, which is commonly considered to be after four times the longest time constant in the system. That corresponds to reducing the influence of transient terms to 1.8% of their original value. Three time constants corresponding to about 5% remaining. For the example system in Equation (14), the 5% condition refers to about 0.33 seconds, and 0.22 seconds has 13.53% remaining.

Consider an ILC problem using system (14), with a sample rate changed to 100 Hz, and consider initially a 21 by 21 matrix $P$, which is reduced to 20 time steps by deleting the first time step from consideration. Because of the faster sample rate, a 4 by 4 block of gains in the upper left corner is adjusting in Approach 1, instead of a 2 by 2 block. For Approach 2 the same 5 by 5 blocks are adjusted. The optimization routine for Approach 2 took longer than before, and for Approach 1 took much longer. To make a fair comparison, the largest singular value in each case was adjusted to be the same, 0.9577, as shown in Tables 7 and 8.

Table 7 First and Last Six Singular Values of $(I - P_1 F_{o1})$ from Large to Small

| Order | $\sigma_1$ | $\sigma_2$ | $\sigma_3$ | $\sigma_4$ | $\sigma_5$ | $\sigma_6$ |
|---|---|---|---|---|---|---|
| Singular value | 0.9577 | 0.1463 | $3.0346e^{-5}$ | $8.2402e^{-6}$ | $7.6463e^{-6}$ | $7.0774e^{-6}$ |
| Order | $\sigma_{15}$ | $\sigma_{16}$ | $\sigma_{17}$ | $\sigma_{18}$ | $\sigma_{19}$ | $\sigma_{20}$ |
| Singular value | $1.7913e^{-6}$ | $1.4988e^{-6}$ | $1.2277e^{-6}$ | $9.0680e^{-7}$ | $4.4318e^{-7}$ | $9.6643e^{-8}$ |

Table 8 First and Last Six Singular Values of $(I - P_1 P_{oc1}^{-1})$ from Large to Small

| Order | $\sigma_1$ | $\sigma_2$ | $\sigma_3$ | $\sigma_4$ | $\sigma_5$ | $\sigma_6$ |
|---|---|---|---|---|---|---|
| Singular value | 0.9577 | 0.9348 | 0.9321 | 0.9247 | 0.9198 | 0.9050 |
| Order | $\sigma_{15}$ | $\sigma_{16}$ | $\sigma_{17}$ | $\sigma_{18}$ | $\sigma_{19}$ | $\sigma_{20}$ |
| Singular value | 0.5687 | 0.4120 | 0.3683 | 0.1307 | 0.0750 | 0.0110 |

Figures 8 and 9 study the learning process a one second period trajectory as in Figures 3 and 4, sampling at 100 Hz, but the desired trajectory is limited to the first 20 time steps. In this case Approach 2 using the circulant matrix appears to have little tracking error during the first ILC iteration, but the Approach 1 result presented in the right of the figure shows very substantial error, seen in Figure 8. Figure 9 presents the corresponding results after 10 iterations for learning, and Approach 2 still has substantial tracking error. It takes more than 40 iterations for the error to become small to graphical accuracy. It appears that Approach 2 using the circulant matrix has substantial advantage learning during the transient phase.

### INVESTIGATING HOW THE ADJUSTED GAINS IN EACH APPROACH MESH WITH THE REMAING GAINS IN THE LEARNING PROCESS

Figures 8 and 9 offer an opportunity to examine how the 4 by 4 set of gains for Approach 1, and the 5 by 5 set of gains for Approach 2 either phase into the remaining gains, or fail to smoothly phase into the remaining gains in the ILC learning process. In Approach 2, the gains in the first 5 time steps (after the unaddressed first time step) are being adjusted, which corresponds to time 0.02 to 0.06. And one might think that after that time one might see some change in the learning. But the learning after 1 and after 10 iterations seem to convert smoothly from the adjusted gains to the unadjusted gains. Note that there are 5 additional adjusted gains at the end of the time interval, and no sudden changes are observed there either. Approach 1 is adjusting the first 4 gains, and one might expect some issue of converting after the 0.02 to 0.05 time interval. It takes many iterations for the output to have what might be considered a smooth transition from adjusted gains to unadjusted gains.

**ROBUSTNESS TO MODEL ERROR**

We examine Approaches 1 and 2 to see if one is significantly better than the other with respect to robustness to model parameter error. Consider robustness to the values of the 3 parameters in the 3$^{rd}$ order system of Equation (14), whose nominal values are $a = 8.8; \omega_0 = 37; \xi = 0.5$. Both the spectral radius stability boundary and the maximum singular value sufficient condition giving monotonic error decay are studied as a function of deviations from nominal for each parameter. Results are presented for each stability criterion and for each parameter in Figure 10. The corresponding ranges of each parameter for error in either direction to satisfy both criteria are presented in Tables 9 and 10. A 50 Hz sample rate is used, with the deleted matrices having 51 rows and 50 columns, sample computations are made for each parameter varying from 1% to 300% of their nominal values. We note that there is rather little difference between the robustness of the two learning gain matrices $F_{o1}$ and $P_{oc1}^{-1}$.

Table 9 Robustness Comparison with Different Parameters Changing ($\sigma_m$)

|  | $a$ | $\omega_0$ | $\xi$ |
|---|---|---|---|
| Monotonic convergence ($P_{oc1}^{-1}, \sigma_m < 1$) | <143% | > 47% and <119% | > 51% |
| Monotonic convergence ($F_{o1}, \sigma_m < 1$) | <159% | > 52% and <130% | > 52% |

Table 10 Robustness Comparison with Different Parameters Changing ($\lambda_m$)

|  | $a$ | $\omega_0$ | $\xi$ |
|---|---|---|---|
| Zero error convergence ($P_{oc1}^{-1}, \lambda_m < 1$) | <187% | > 32% and <136% | > 48% |
| Zero error convergence ($F_{o1}, \lambda_m < 1$) | <176% | <136% | > 7% |

**DISCUSSION AND CONCLUSIONS**

This paper presents how one can use frequency response design methods to design Iterative Learning Controllers. There are two versions available, use of a Finite Impulse Response filter to fit the inverse of the system frequency response, and use of the circulant matrix that is known to give the steady state frequency response, but only for the discrete frequencies that can be seen in the number of time steps of data available. The purpose of this paper is to compare these two approaches.

These laws are based on inverting the frequency response of the system to update each ILC iteration. This bypasses any attempt to invert the discrete time transfer function which usually results in unstable commands. But it is natural to think that inverting steady state frequency response could make a good control update, but only for parts of the trajectory each run that are beyond the transient effects of the initial conditions. To have this approach address the error during the transients in every iteration in ILC seems like wishful thinking. And more extreme, one would not expect the inverse steady state frequency response to be successful when the whole finite time trajectory is within the system transients, or settling time. However, it appears here that by adjusting a relatively small number of initial frequency response gains in the learning gain matrix, a number much smaller than the number of time steps in the transients, the approach produces convergent ILC laws. In fact, these laws can be tuned to give much faster convergence than one usually gets from routine time domain control laws. And it can do so with monotonic decay of the Euclidean norm of the error. These two approaches to frequency response based ILC design, represent a new family of ILC laws.

In comparing the two approaches, there appears to be very little difference in performance. After adjustment of learning gains to make comparable maximum singular values for a fair comparison, the RMS error reduction from run to run exhibits little difference in performance. Each approach has separate kinds of difficulties in representing the steady state frequency response for the early time steps, and also final time steps in the case of circulant matrices. Perhaps surprisingly, the adjustment of a few gains is seen to handle these difficulties well. Also, no significant difference between the two approaches is seen when studying their robustness to model parameter error.

The conclusion is that there seems to be no significant difference in performance of the two approaches. Perhaps the main difference is that there is more computation needed to make the adjustments needed to gains in the circulant based method than in the Finite Impulse Response based method.

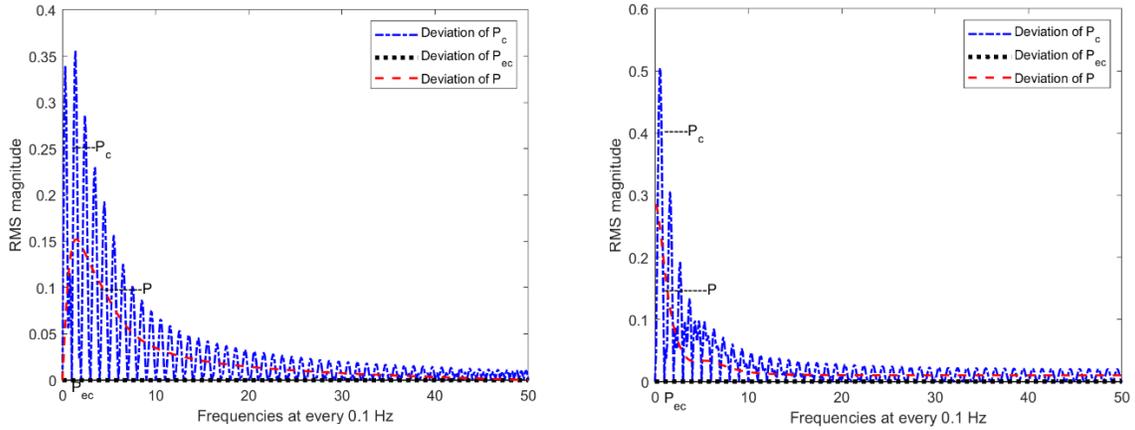

Figure 1. RMS deviation of frequency response of $P_c$, $P_{ec}$, $P$ compared to discrete time transfer function response with system equation (14), input $\sin(wt)$ (left) and $\cos(wt)$ (right)

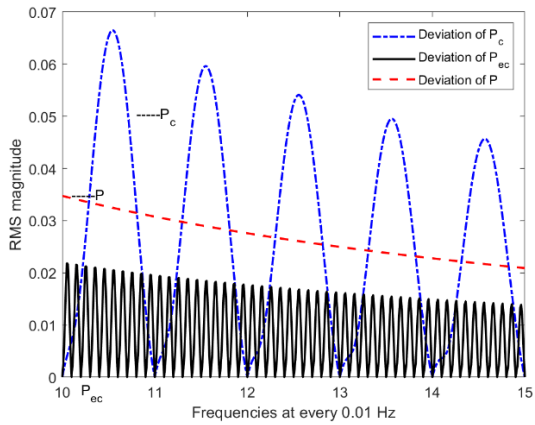
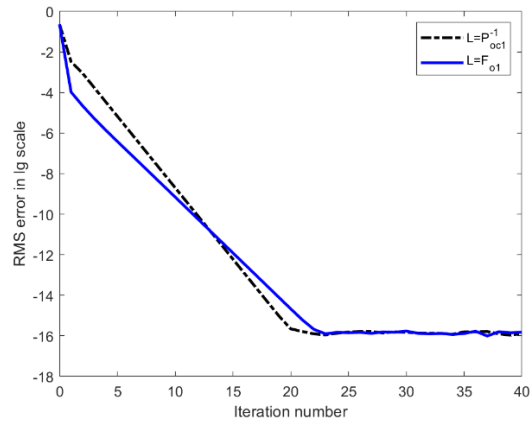

Figure 2. Detail of $\sin(\omega t)$ of Figure 1, showing deviation of $P_c$, $P_{ec}$, $P$ with trajectory extended from 1 second to 10 seconds

Figure 3. RMS error using $L = P_{oc1}^{-1}$ and $L = F_{o1}$ versus iteration. $y^*(t) = \pi(5t^3 + 7.5t^4 + 3t^5)$ and initial run applies $y^*(t)$

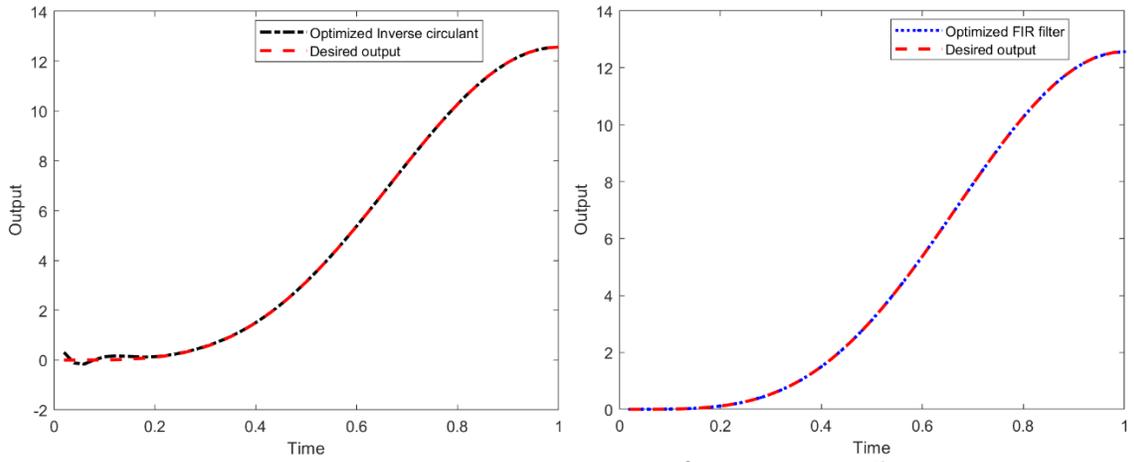

Figure 4. Half period of desired trajectory $y^*(t) = \pi(1 - \cos(wt))^2$ and output after $1^{st}$ ILC iteration using $L = P_{oc1}^{-1}$ (left) and $L = F_{o1}$ (right) with sample rate 50Hz and trajectory full period 2 seconds

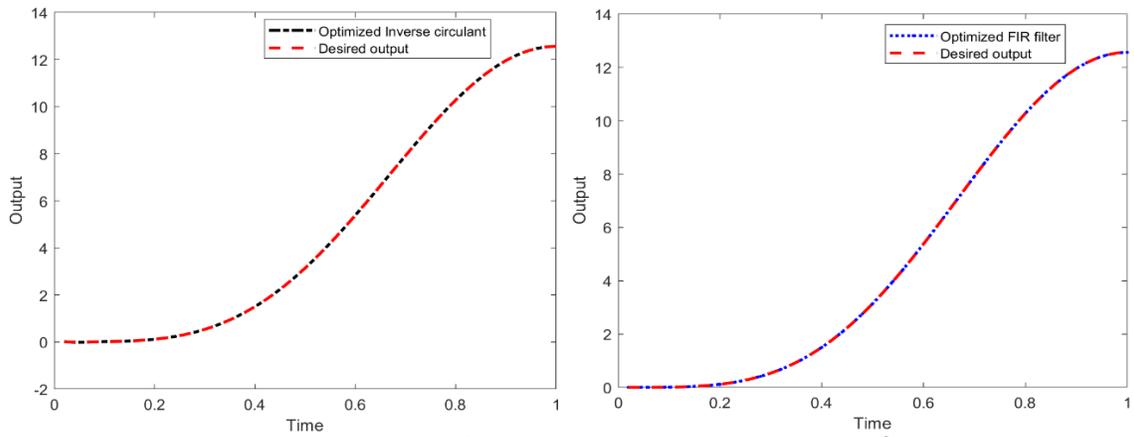

Figure 5. Half period of desired trajectory $y^*(t) = \pi(1 - \cos(wt))^2$ and output after $3^{rd}$ ILC iteration using $L = P_{oc1}^{-1}$ (left) and $L = F_{o1}$ (right) with sample rate 50Hz and trajectory full period 2 seconds

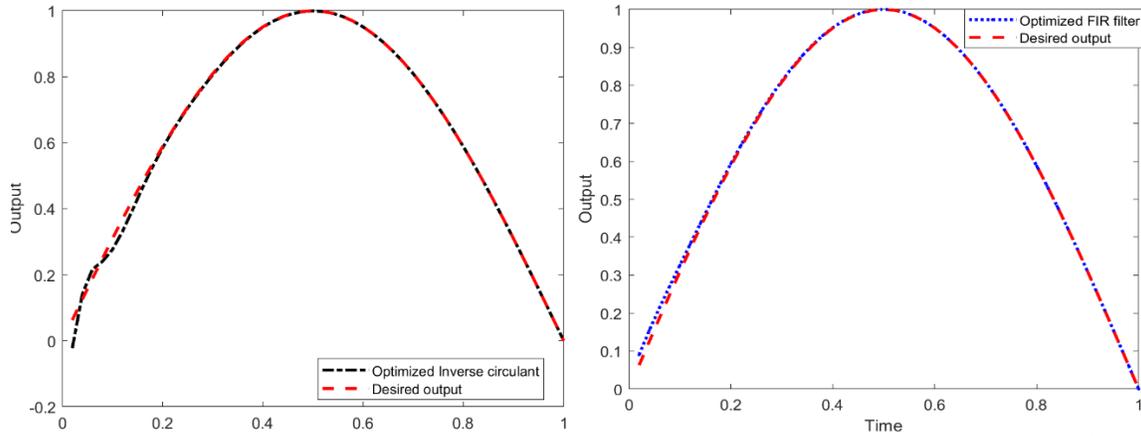

Figure 6. Half period of desired trajectory $y^*(t) = \sin(\omega t)$ and output after $1^{st}$ ILC iteration using $L = P_{oc1}^{-1}$ (left) and $L = F_{o1}$ (right) with sample rate 50Hz and trajectory full period 2 seconds

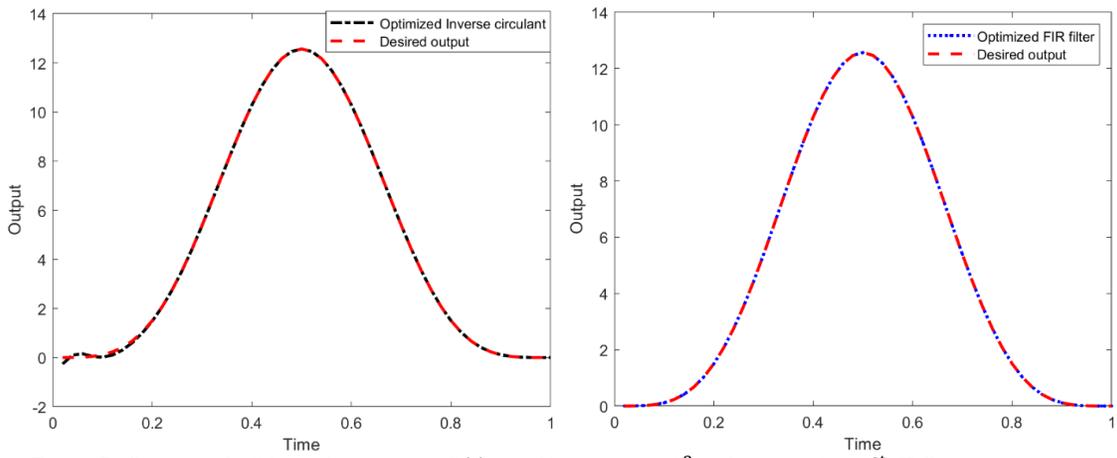

Figure 7. One period of desired trajectory $y^*(t) = \pi(1 - \cos(wt))^2$ and output after $1^{st}$ ILC iteration using $L = P_{oc1}^{-1}$ (left) and $L = F_{o1}$ (right) with sample rate 50Hz and trajectory full period 1 second

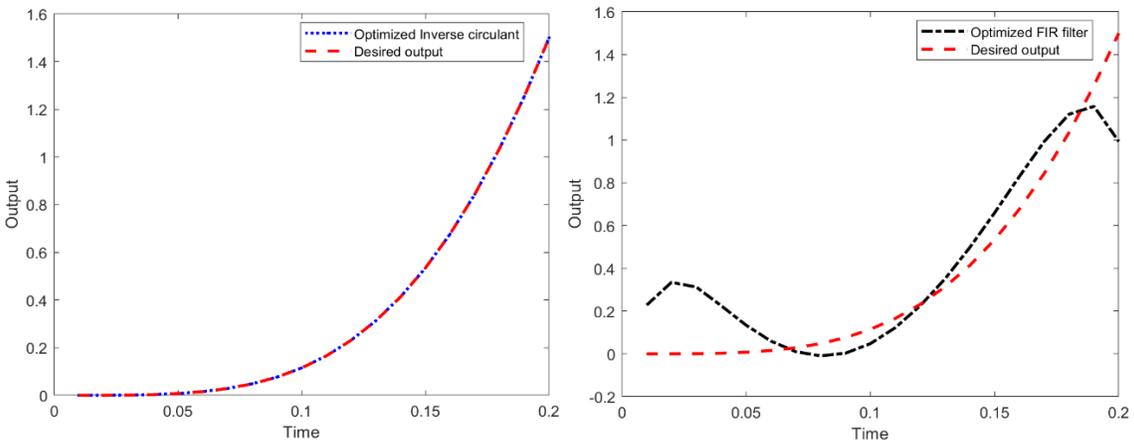

Figure 8. $\frac{1}{5}$ period of desired trajectory $y^*(t) = \pi(1 - \cos(wt))^2$ and output after $1^{st}$ ILC iteration using $L = P_{oc1}^{-1}$ (left) and $L = F_{o1}$ (right) with sample rate 100Hz and trajectory full period 1 second

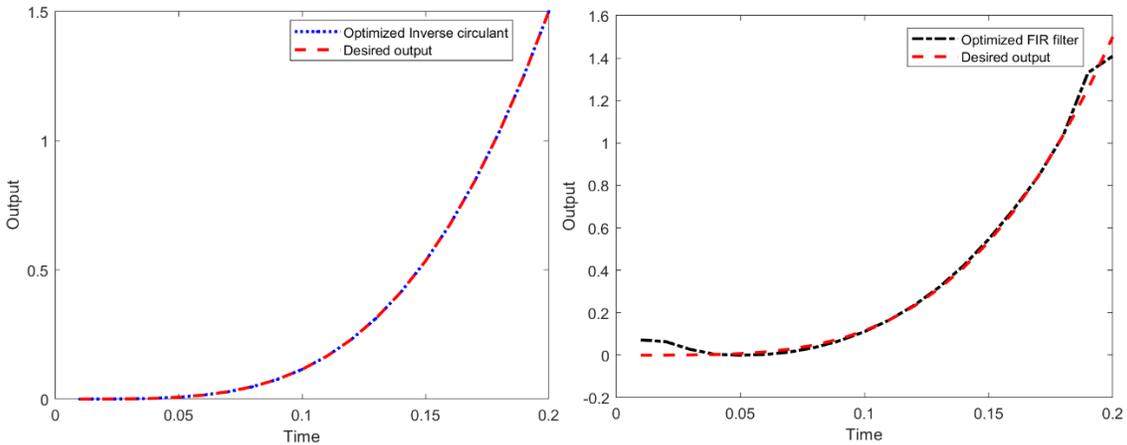

Figure 9. $\frac{1}{5}$ period of desired trajectory $y^*(t) = \pi(1 - \cos(wt))^2$ and output after $10^{th}$ ILC iteration using $L = P_{oc1}^{-1}$ (left) and $L = F_{o1}$ (right) with sample rate 100Hz and trajectory full period 1 second

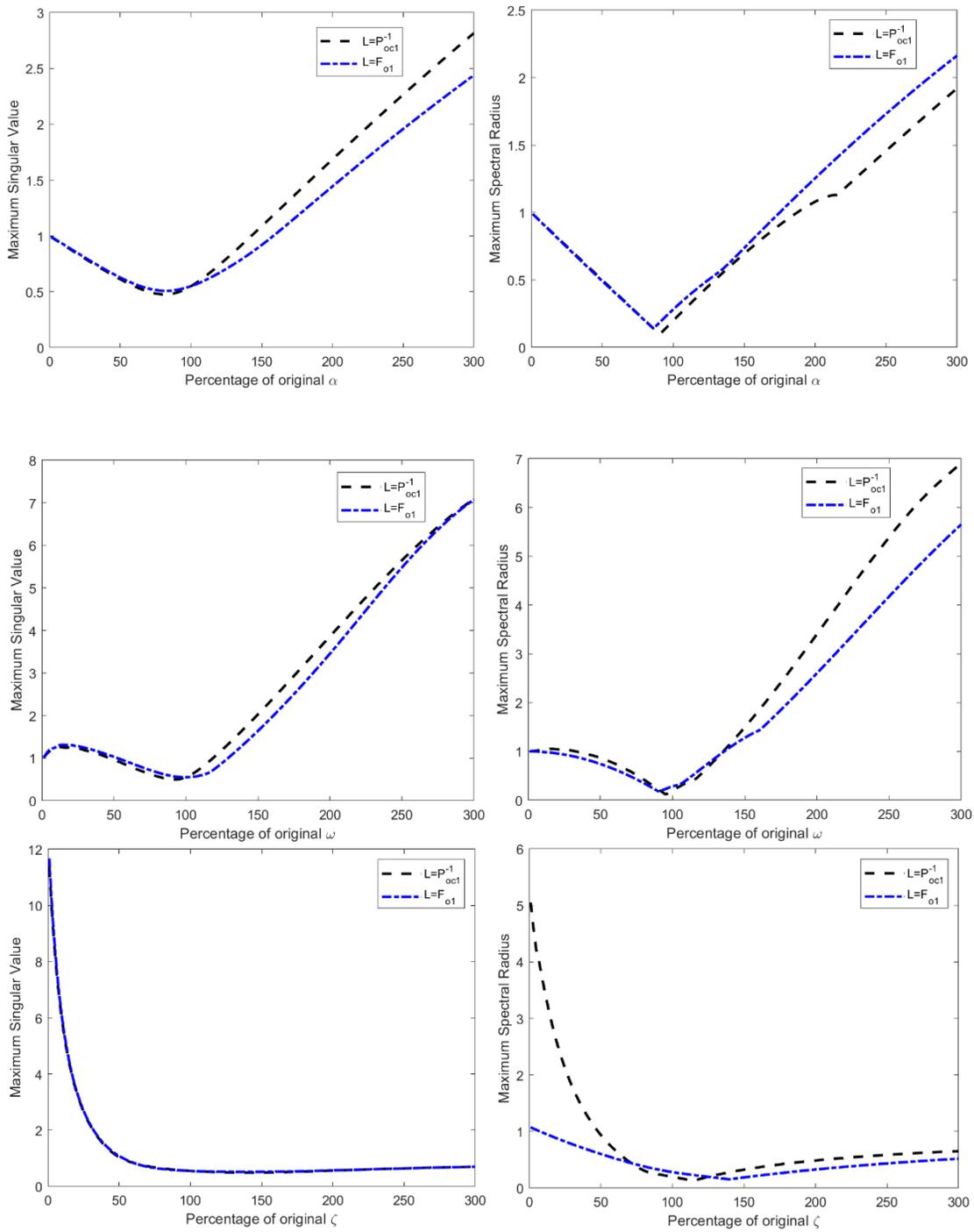

Figure 10. Maximum singular value (left), spectral radius (right), of $(I + P_1 L)$ for variation of $\alpha$, $\omega$, and $\zeta$ (top, middle, bottom) from nominal values (100) and percent increase/decrease.